\documentclass[letterpaper, 10 pt, conference]{ieeeconf}  

\IEEEoverridecommandlockouts                              

\usepackage{graphics} 
\usepackage{epsfig} 
\usepackage{mathptmx} 
\usepackage{times} 
\usepackage{amsmath} 
\usepackage{amssymb}  
\usepackage{xcolor}
\usepackage{multirow}
\usepackage{subfigure}
\usepackage{caption}
\usepackage{subcaption}  

\title{\LARGE \bf
Optimal Design of Experiment for Electrochemical Parameter Identification of Li-ion Battery via Deep Reinforcement Learning
}

\author{Mehmet Fatih Ozkan, Samuel Filgueira da Silva, Faissal El Idrissi, Prashanth Ramesh
and Marcello Canova
\thanks{Department of Mechanical and Aerospace Engineering, Center for Automotive Research, The Ohio State University, Columbus, OH, USA. Emails: 
        {\tt\small ozkan.25@osu.edu, filgueiradasilva.1@osu.edu, canova.1@osu.edu}}%
}

\begin{document}

\maketitle
\thispagestyle{empty}
\pagestyle{empty}

\begin{abstract}
Accurate parameter estimation in electrochemical battery models is essential for monitoring and assessing the performance of lithium-ion batteries (LiBs). This paper presents a novel approach that combines deep reinforcement learning (DRL) with an optimal experimental design (OED) framework to identify key electrochemical parameters of LiB cell models. The proposed method utilizes the twin delayed deep deterministic policy gradient (TD3) algorithm to optimize input excitation, thereby increasing the sensitivity of the system’s response to electrochemical parameters. The performance of this DRL-based approach is evaluated against a nonlinear model predictive control (NMPC) method and conventional tests. Results indicate that the DRL-based method provides superior information content, reflected in higher Fisher information (FI) values and lower parameter estimation errors compared to the NMPC design and conventional test practices. Additionally, the DRL approach offers a substantial reduction in experimental
time and computational resources.
\end{abstract}
\section{Introduction}
The accurate estimation of the parameters in electrochemical models of li-ion battery (LiB) cells is crucial for cell and pack design, system monitoring, diagnostics, and control \cite{anderson2022}. Parameter estimation involves determining parameter values by fitting a model to measure input-output data using specific algorithms. This task is quite challenging due to the nonlinearity and interdependence of the processes involved, making it difficult to distinguish the underlying phenomena attributed to discrepancies between the model predictions and experimental data \cite{pozzi2020,chen2020development}. Consequently, the calibration process necessitates a deep understanding of the model, often requiring sophisticated optimization techniques and extensive experimental validation.\par 
While research has extensively focused on parameter estimation algorithms, the significance of data quality has been relatively underexplored \cite{Lin2019review}. Data quality profoundly affects parameter estimation such that the data lacking information about the target parameter can limit estimation accuracy, while data with undesirable structures may amplify errors arising from system uncertainties \cite{Lai2011}. Recent efforts have focused on designing input excitations to generate optimal data for estimation. In control theory and system identification, this input design is treated as a form of optimal experimental design (OED), aimed at maximizing metrics related to data quality, such as parameter sensitivity or Fisher information (FI). FI is frequently employed in input optimization for battery parameter estimation \cite{park2018optimal,lai2020optimization,huan22vppc,huang2023reinforcement,huang2023excitation}, and serves as a valuable metric for designing optimal experiments that maximize the information available for estimating parameters of interest in a battery system. 

Optimization in OEDs can be computationally demanding, particularly when iterative calculations and the inversion of design criteria or system dynamics are required. Traditional gradient-driven optimization methods often struggle with achieving global optimality, especially in non-convex and high-dimensional spaces. Recent studies address these challenges by framing input generation as a Markov Decision Process (MDP), with dynamics linked to the battery's physical states, and applying a Reinforcement Learning (RL) approach to develop control policies that optimize design criteria, such as maximizing FI over the specified horizon \cite{huan22vppc,huang2023reinforcement,huang2023excitation}. These studies demonstrate that RL-based methods can yield high FI and lower parameter estimation errors for the target parameters compared to the open-loop conventional nonlinear optimization approach.\par 
While these existing works have demonstrated the effectiveness of OEDs in LiBs, there are several limitations. First, RL-based methods have yet to be compared with more advanced closed-loop controllers like Model Predictive Control (MPC), which handles constraints and optimizes performance over time. A comparison with MPC could provide valuable insights into the advantages of RL-based approaches in this context. Second, these studies primarily relied on Q-learning, which is effective in certain scenarios but limited in managing the exploration-exploitation tradeoff and scalability in high-dimensional spaces \cite{DRL}, which are often found in LiB models. More advanced deep RL (DRL) algorithms like Twin Delayed Deep Deterministic Policy Gradient (TD3) \cite{fujimoto2018addressing} have shown promise in LiB applications owing to their applicability to difficult optimization problems \cite{Chun2024}. Lastly, the focus of these existing studies was primarily on optimizing current excitation for identifying a limited number of electrochemical parameters such as electrode diffusion coefficient and active material volume fraction \cite{huang2023reinforcement}, \cite{huang2023excitation}. The objective of this study is to leverage the DRL algorithm to optimize input excitation for parameter identification in electrochemical models of LiB cells and to test its performance and robustness against the nonlinear MPC (NMPC) design and conventional test procedures.


\section{Overview of Li-ion Cell Model Equations}

\subsection{Battery Model}
This study uses the electrochemical equivalent circuit model (E-ECM) of a 30 Ah NMC-graphite cell as a computationally efficient model to reduce the computation time required for OEDs. The E-ECM linearizes and reduces the order of the governing equations of the Extended Single-Particle Model (ESPM), yielding a mathematical form that enables faster computations while maintaining the connection between model parameters and the physical processes from the original equations. The key constitutive equations of the E-ECM are summarized for clarity, with a comprehensive description of the assumptions and derivation framework available in \cite{seals2022physics}. \par
The terminal voltage $V$ of the E-ECM model takes the form:
\begin{equation} \label{eq:1}
\begin{split}
V(t) = U_p(c_{se,p},t) - U_n(c_{se,n},t) - \eta_p(c_{se,p},t) \\ + \eta_n(c_{se,n},t) + \phi_{diff}(t) + \phi_{ion}(t) - I(t)R_c
 \end{split}
\end{equation}
which is composed of seven terms. The first two terms relate to the open circuit potential (OCP) $U_i$, which is a function of the electrode’s surface concentration $c_{se,i}$. The third and fourth terms represent the kinetic overpotential at the solid-electrolyte interface (SEI) $\eta_{i}$, which is given by the Butler-Volmer equation. The fifth and sixth terms represents overpotentials in the electrolyte due to diffusion  $\phi_{diff}$ and ionic conductivity $\phi_{ion}$, respectively. Finally, $R_c$ accounts for lumped resistance considering the current collectors and contact resistance. The subscripts $i=p$ and $i=n$ refer to the positive and negative electrodes, respectively.

\subsection{Parameter Sensitivity}
In physics-based electrochemical models, the local sensitivity of the terminal voltage to each parameter is quantified by calculating the first-order partial derivative with respect to that parameter at each time instant. The sensitivity of parameter $\theta$ measures how changes in $\theta$ affect the variation in the output. The sensitivity can be used to define the quality of the data via the FI \cite{lehmann2006theory}, which measures the amount of information about the parameter $\theta$ contained in the observed output $y_1, \ldots, y_k, \ldots, y_N$ with $N$ data points. In the presence of
i.i.d. Gaussian measurement noise, the FI can be calculated as \cite{scharf1993geometry}:
\begin{equation}
    \text{FI} = \frac{1}{\sigma_y^2} \sum_{k=1}^N \left( \frac{\partial y_k}{\partial \theta} \right)^2
\end{equation}
where $\sigma_y^2$ represents the variance of the measurement error in the output and $\frac{\partial y_k}{\partial \theta}$ denotes the $y_k$ sensitivity of parameter $\theta$ at data point $k$. The inverse of the FI provides the Cramér-Rao bound, which establishes the lower bound for the variance of the estimation error of an unbiased estimator of $\theta$:
\begin{equation}
    \sigma^2(\hat{\theta}) \geq \text{FI}^{-1}
\end{equation}
\par In this work, the OED procedure is applied to enhance the estimation of the anode and cathode rate constants ($k_n$, $k_p$) which are critical parameters influencing the dynamic response of the electrochemical models,  as reported for instance in \cite{dangwal2021parameter}. The sensitivities of such parameters can be expressed as:
\textcolor{black}{\begin{equation} \label{eq:2}
\begin{split}
\frac{\partial V(t)}{\partial k_i} = \frac{\partial U_p(t)}{\partial k_i} - \frac{\partial \eta_p(t)}{\partial k_i} - \frac{\partial U_n(t)}{\partial k_i} +  \frac{\partial \eta_n(t)}{\partial k_i} - \\ \frac{\partial \phi_e(t)}{\partial k_i}  - \frac{\partial R_c(t)}{\partial k_i}
 \end{split}
\end{equation}}
\noindent After canceling terms which are not the function of the rate constants this leads to:
\textcolor{black}{\begin{equation} \label{eq:3}
\frac{\partial V(t)}{\partial k_p} =   -\frac{\partial \eta_p(t)}{\partial i_{0p}}\frac{\partial i_{0p}(t)}{\partial k_p},\ \ \frac{\partial V(t)}{\partial k_n} =   \frac{\partial \eta_n(t)}{\partial i_{0n}} \frac{\partial i_{0n}(t)}{\partial k_n} 
\end{equation}}
\noindent where the kinetic overpotential $\eta_{i}$ and the exchange current density $i_{0i}$ are expressed as:
\setlength{\belowdisplayskip}{5pt} \setlength{\belowdisplayshortskip}{5pt}
\textcolor{black}{\begin{equation} \label{eq:5}
\begin{split}
\eta_{i}(t) = \frac{\bar R T_0(-J_iI(t))}{Fi_{0,i}}
 \end{split}
\end{equation}}
\textcolor{black}{\begin{equation} \label{eq:6}
\begin{split}
i_{0i}(t) = exp\left( \left(  \frac{1}{T_{ref}} - \frac{1}{T(t)} \right) \frac{E_{io,i} }{\bar R}   \right) F k_i ~ \times \\  \sqrt{c_{se,i}(c_{max,i} - c_{se,i})c_{e,i}}
 \end{split}
\end{equation}}
\noindent where $J_i$ denotes the intercalation current density, $E_{io,i}$ refers to the activation energy, $c_{max,i}$ represents the maximum surface concentration, $c_{e,i}$ is the electrolyte concentration, and $T_{ref}$ is the reference temperature.

\section{Optimal Experimental Design}
Leveraging the E-ECM and parameter sensitivity analysis, an optimal experiment can be developed by specifying an objective function and then solving the resulting optimization problem. This section will present both the proposed DRL method and the NMPC design for the OED procedure. 
\subsection{Deep Reinforcement Learning}
In this work, an actor-critic DRL method TD3 algorithm is proposed to optimize input excitation in LiBs. TD3, an advanced variant of the Deep Deterministic Policy Gradient (DDPG) algorithm, is selected for its superior stability and performance in continuous action spaces, making it highly suitable for this application \cite{fujimoto2018addressing}. Unlike the model-based approaches, which depends on a discretized E-ECM and requires numerous states to capture system behavior, TD3 is model-free and does not rely on an explicit mathematical model of the battery dynamics, allowing for greater flexibility and easier implementation, especially in real-time scenarios.

In the proposed DRL-based input excitation, the problem of generating the optimal excitation is formulated as an MDP where the dynamics are characterized by specific continuous states ($s$) influenced by the underlying dynamics associated with the E-ECM, specifically cell terminal voltage ($V$) and normalized lithium surface concentration in the cathode ($\overline{c}_{se,p}$). The objective is to determine an optimal control policy, denoted as $\pi(s|a)$, which maximizes the reward function ($r$) and maps the states ($s$) to an 
 action ($a$), which is the electrical current ($I$) in this context. RL learns the policy $\pi(s|a)$ by exploring the state-action space to solve the Bellman Equation. The Bellman Equation calculates the value of the specific states ($s$) by considering the immediate reward of taking a particular action in those states, along with the expected future rewards from the resulting states. This computation helps determine the overall value of being in the given state, providing insights into its significance within a system or decision-making process. The Q-function uses the Bellman equation and evaluates the value of the continuous states ($s$) and action ($a$) under a given policy $\pi(s|a)$. The Q-function formulates the joint desirability of a state and action pair for optimized learning, and is defined as follows:
\begin{equation}
Q\left(s \mid a\right) = \max_{\pi(s \mid a)} \mathbb{E}\left[r_t + \sum_{k=1}^{\infty} \gamma^k r_{t+k} \mid s_{t+1}, a_{t+1}\right]
\end{equation}
where $\gamma$ is a discount factor and $r_{t+k}$ is reward at time step $t+k$. The actor-critic architecture of TD3 involves two networks: the actor network, responsible for learning the policy, and the critic network, which estimates the Q-values. This separation ensures more stable training and learning, particularly in continuous action spaces compared to traditional methods like Q-learning \cite{fujimoto2018addressing}. In this DRL-based design of experiment formulation, the reward function consists of two different terms. The first term maximizes the FI for the considered parameter $FI_{\theta_i}$, while the second term penalizes any constraint violations on the voltage output. The applied single-step reward function is defined as:
\setlength{\belowdisplayskip}{5pt} \setlength{\belowdisplayshortskip}{5pt}
\begin{equation}
r_k =
\begin{cases}
    \left(\frac{\partial V_k}{\partial \theta_i}\right)^2, & \text{normal condition} \\
M, & \text{constraint violation}
\end{cases}
\end{equation} \par
\subsection{Nonlinear Model Predictive Control}
This work introduces an NMPC approach as an alternative strategy for optimizing input excitation in LiBs. The NMPC is chosen due to its established capability to handle nonlinear dynamics and constraints while optimizing a defined objective over a defined prediction horizon. The NMPC approach is formulated using the discretized E-ECM with zero-order hold (ZOH), ensuring an accurate representation of system dynamics in the optimization framework. In this formulation, input vector, $u$, contains electrical current ($I$) and state vector, $x$, contains nine states including cell terminal voltage ($V$), surface concentration ($c_{se,i}$), bulk concentration states ($c_{b1,i}$, $c_{b2,i}$) and diffusion dynamics contribution ($c_{d,i}$) in positive and negative electrodes, respectively. The control objective of the NMPC design is to maximize the FI of the target parameter subject to the system constraints. Therefore, the NMPC framework is designed to minimize the negative of the FI related to the target electrochemical parameter $\theta_i$ over a finite prediction horizon \( H \). The objective function is defined as:
\begin{equation}
\min_{\{u_k\}_{k=0}^{H-1}} \sum_{k=0}^{H-1} -\left( \frac{\partial V_k}{\partial \theta_i} \right)^2
\end{equation}
subject to the nonlinear state dynamics, current and voltage constraints:
\begin{equation}
x_{k+1} = f(x_k, u_k,\Theta), \ I_{min}\leq I_{k}\leq I_{max},\ V_{min}\leq V_{k}\leq V_{max}
\end{equation}
where \( x_k \) and \( u_k \) represent the state and input vector at step \( k \), respectively, $\Theta$ defines the set of parameters and \( f(x_k, u_k,\Theta) \) represents the discretized dynamics of the LiB cell as described by the E-ECM.


\section{Results and Discussion}
We analyzed and compared the current excitations optimized for estimating the \(k_p\) and \(k_n\) parameters using the proposed DRL-based approach against the NMPC approach and conventional tests including 1C constant current (CC) discharging, relaxation current interrupt discharge (RCID) \cite{hu2011electro} and experimental drive cycle tests from \cite{pi2024parameter}. The optimization tasks are designed to maximize the FI of the target parameters in both DRL and NMPC designs, with the battery starting at a SoC of 100\% over an 1800-second time horizon. For the DRL-based approach, each training episode begins at 100\% SoC and concludes after 1800 steps have elapsed or terminates early due to constraint violations. A step size of 1 second was used, and training was conducted with episode counts ranging from 1,000 to 10,000 episodes.

Table \ref{table:setting} presents information on the operating conditions of the simulated battery system, the penalty for constraint violation, and the hyperparameters of the TD3 algorithm. RL toolbox in MATLAB/Simulink is used to train and test the TD3 algorithm. In the NMPC design, the prediction horizon is set to 20 s with 1 s sample time. The same input and output constraints in the DRL settings are applied to the NMPC design for a fair comparison and the sequential quadratic programming (SQP) algorithm in MATLAB is used to solve the NMPC approach. 

\begin{table}[h!]
\centering
\caption{TD3 Model Parameters and Settings}
\label{table:setting}
\begin{tabular}{|l|l|}
\hline
\textbf{Parameter} & \textbf{Value} \\ \hline
Current Input Range & \([-150 \text{ A}, 150 \text{ A}]\) \\ \hline
Voltage Output Range & \([2.8 \text{ V}, 4.2 \text{ V}]\) \\ \hline
Penalty for Constraint Violation, \(M\) & -5 \\ \hline
Simulation Length  and Time step & 1800 s and 1 s\\ \hline
Discount Factor, $\gamma$ & 0.99 \\ \hline
Experience Buffer Length & \(2 \times 10^7\) \\ \hline
Mini-Batch Size & 256 \\ \hline
Maximum Training Episodes & 10000 \\ \hline
Actor and Critic Learning Rate & $10^{-4}$ and $10^{-3}$\\ \hline
Exploration Noise and Model & 9 A and Ornstein-Uhlenbeck\\ \hline
Optimizer & Adam \\ \hline
\end{tabular}
\label{table:model_params}
\end{table}
\subsection{Parameter Sensitivity Analysis}
\begin{figure}[!h]
    \centering
    \includegraphics[angle=0, scale=0.45]{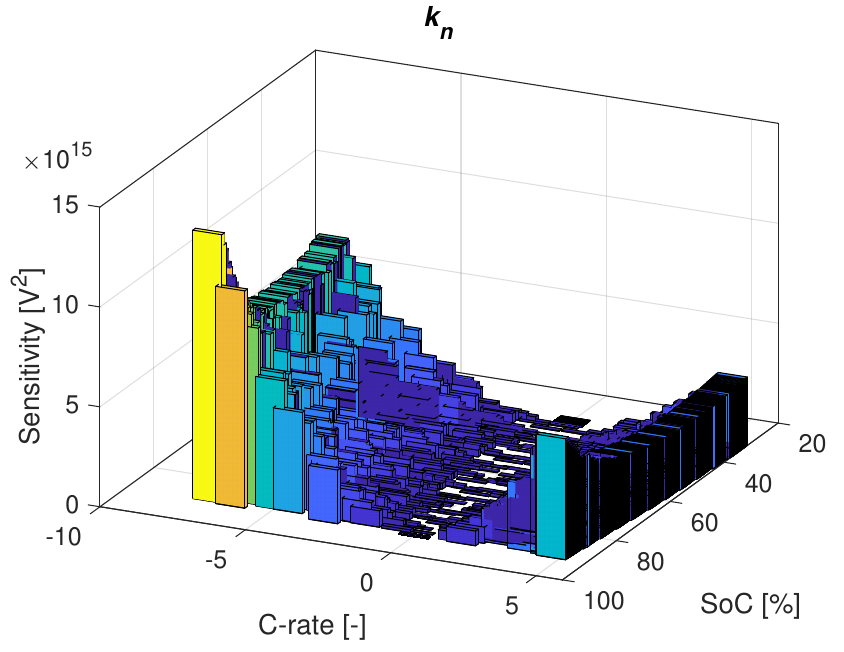}
    \includegraphics[angle=0, scale=0.45]{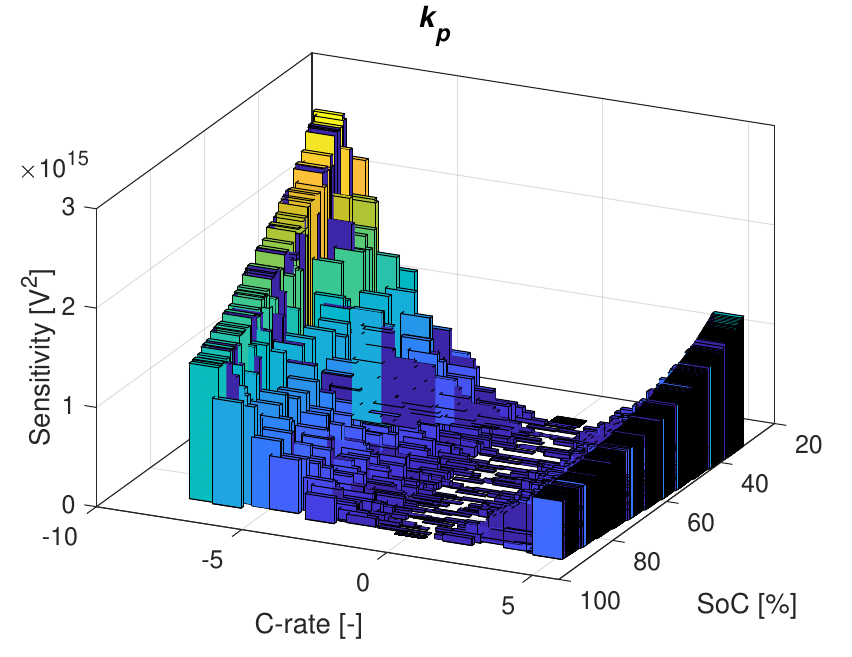}
    \caption{Sensitivity (squared) of voltage output on $k_n$ and $k_p$ parameters with respect to the C-rates and SoC levels under drive cycle test.}
   \label{fig:sensitivity}
\end{figure}

A sensitivity analysis is performed using the experimental drive cycle test from \cite{pi2024parameter} to provide foundational insights into the role of the rate constant parameters in the overall battery behavior. The sensitivity profiles with respect to the C-rates and SoC levels are plotted in Fig. \ref{fig:sensitivity}.\par


Fig. \ref{fig:sensitivity} reveals that the voltage is highly sensitive to variations in $k_n$, particularly at high SoC levels. This sensitivity can be attributed to the increased surface concentration of lithium ions at the anode as the SoC approaches its upper limits. At high SoC, the anode operates near its maximum capacity, where any deviation in the reaction kinetics, as dictated by the rate constant, has a pronounced effect on the voltage. The effect of this sensitivity is further amplified at higher C-rates, where faster reaction kinetics are required to sustain the higher current. 

On the other hand, the sensitivity of the voltage to the $k_p$ exhibits a contrasting trend, with high sensitivity observed at low SoC levels. This increased sensitivity at low SoC can be attributed to the depletion of active lithium material in the cathode as it nears full discharge. At low SoC, the reaction kinetics depend more on the rate constant to sustain the voltage, as the remaining lithium ions become more difficult to intercalate. Similar to the anode rate constant, the impact of the C-rate is significant here. At higher C-rates, faster reaction kinetics are required, which amplifies the effects of diffusion limitations and further increases the voltage sensitivity to changes in the cathode rate constant. These factors overall make the voltage particularly vulnerable to changes in $k_n$ and $k_p$ at high and low SoC, respectively, especially under high C-rates.\par
\subsection{Analysis of Optimal Experimental Designs}
\begin{figure}[h!]
    \centering
    \subfigure{{\includegraphics[angle=0, scale=0.50]{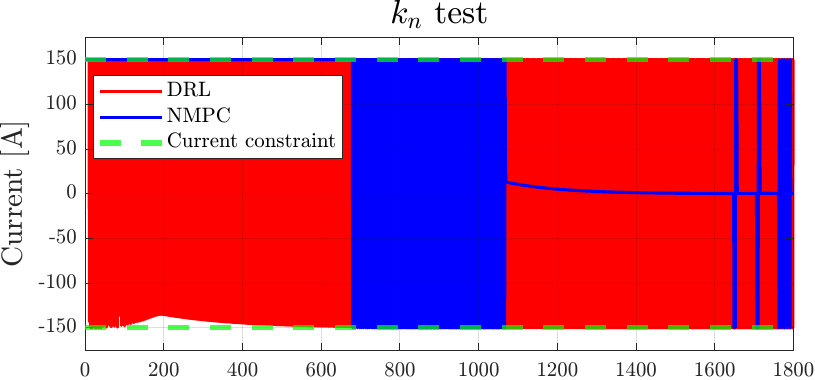}}}
    \subfigure{{\includegraphics[angle=0, scale=0.50]{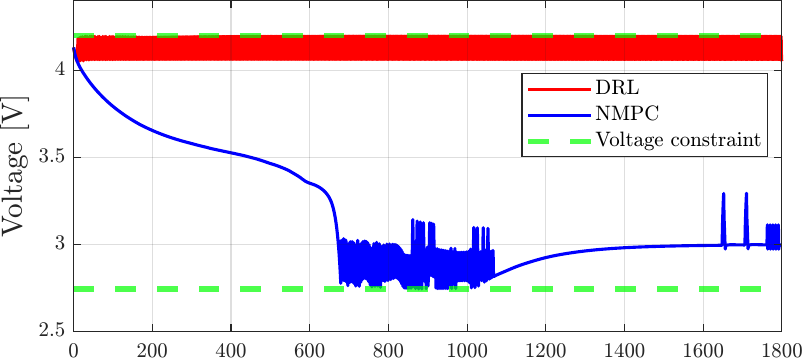}}}
    \subfigure{{\includegraphics[angle=0, scale=0.50]{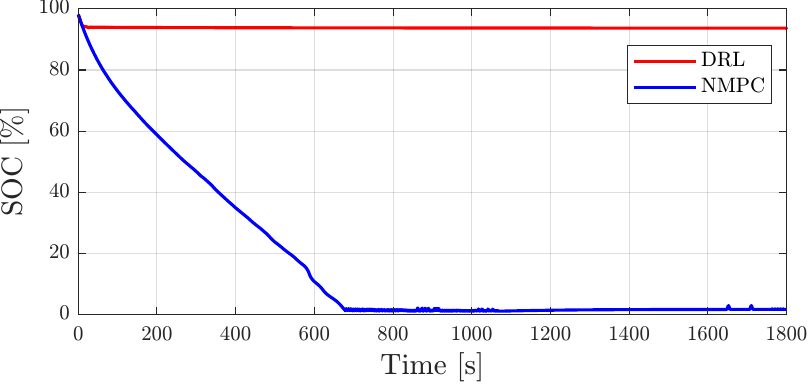}}}
    \caption{ Comparison of DRL and NMPC based design of experiments for $k_n$ parameter.}
    \label{fig:DRLxNMPC_kn}
\end{figure}

This section evaluates the optimization results of the proposed DRL and NMPC approaches. The optimized current excitation profiles are analyzed and the FI contained in the generated profiles are compared to assess the effectiveness of each approach. As shown in the optimization results in Fig. \ref{fig:DRLxNMPC_kn}, the DRL-trained policy frequently excites the battery with high C-rate discharging and charging at high SoC for $k_n$, while the NMPC approach depletes the battery to the low SoC region and performs high C-rate discharging and charging. This behavior highlights the distinct results obtained with each approach and relates closely to the sensitivity analysis of the rate constants mentioned in the previous section.\par
\begin{figure}[h!]
    \centering
    \subfigure{{\includegraphics[angle=0, scale=0.50]{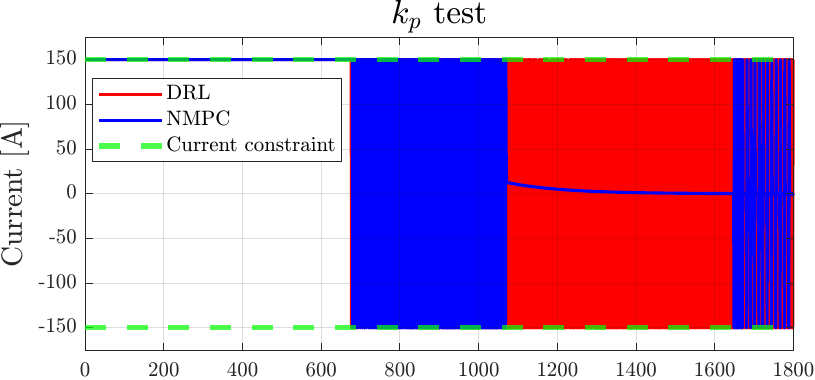}}}
    \subfigure{{\includegraphics[angle=0, scale=0.50]{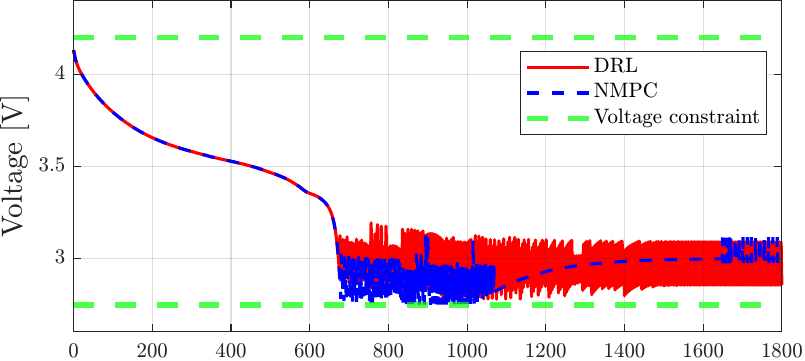}}}
    \subfigure{{\includegraphics[angle=0, scale=0.50]{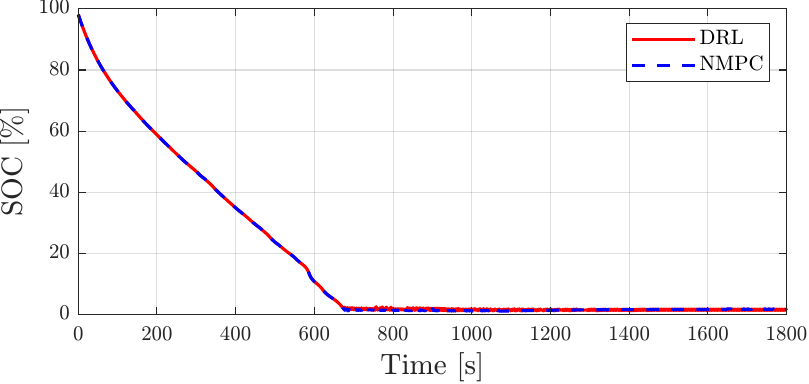}}}
    \caption{ Comparison of DRL and NMPC based design of experiments for $k_p$ parameter.}
    \label{fig:DRLxNMPC_kp}
\end{figure}

The DRL-trained policy leverages the high sensitivity of $k_n$ at high SoC. By focusing on this region, the DRL approach optimizes current excitation to enhance parameter estimation accuracy, capitalizing on the pronounced effects of $k_n$ on voltage at high C-rate conditions. This tailored excitation design allows the DRL policy to effectively navigate the dynamics of $k_n$, making it particularly effective for high SoC operations. Conversely, NMPC approach does not leverage the high sensitivity of $k_n$ as effectively as the DRL-trained policy does in the high SoC region and targets the low SoC region, where the sensitivity of $k_n$ is lower compared to the high SoC region. The primary reason for this difference is that the NMPC with limited preview information lacks prior knowledge of the maximum sensitivity region at high SoC. As a result, it explores the SoC space by discharging the cell and applies high C-rate discharging and charging once it reaches low SoC, favoring $k_n$ sensitivity in that region. In contrast, the DRL approach leverages prior knowledge of the maximum sensitivity at high SoC gained during training, applying high C-rate discharging and charging in the high SoC region to favor $k_n$ sensitivity. Both approaches, however, show a similar pattern in optimizing $k_p$. As seen in Fig.~\ref{fig:DRLxNMPC_kp}, both the DRL-trained policy and the NMPC approach deplete the battery to a low SoC region and perform high C-rate discharging and charging, effectively utilizing the increased sensitivity of $k_p$ at low SoC to enhance parameter estimation. It is also observed that the DRL-trained policy applies high C-rate charging and discharging pulses more frequently than the NMPC approach, resulting in higher FI.\par

These findings are further supported by quantitative comparison in Table \ref{table:1}, where the DRL approach achieves significantly higher average FI for $k_n$ and $k_p$ compared to the NMPC design. Both qualitative and quantitative results overall demonstrate that the proposed DRL-based approach is more capable of finding the optimal excitation for $k_n$ and $k_p$ parameters compared to the NMPC approach.

\begin{table*}[h!]
\centering
\caption{Parameter estimation results.}
\label{table:1}
\begin{tabular}{ccccccc} 
\hline
\multirow{2}{*}{\textbf{Current profile}} & \multicolumn{2}{c}{\textbf{Average Fisher Information [V$^2$]}} & \multicolumn{2}{c}{\textbf{Parameter Estimation Error (Median) [\%]}} & \multirow{2}{*}{\textbf{Experiment Length [s]}} \\ 
& \textbf{$k_p$} & \textbf{$k_n$} & \textbf{$k_p$} & \textbf{$k_n$} & \\
\hline
1C Discharge Test @ 25$^{\circ}$C    &  2.20e+10    & 1.00e+11 &  1.903    & 1.492 & 3800 \\
RCID Test @ 45$^{\circ}$C        &   5.76e+10    & 6.71e+09 &  1.072    & 0.721 & 102212 \\
Standard Drive Cycle   @ 45$^{\circ}$C     &    1.52e+10   & 6.56e+10 &  2.381   & 2.190 & 1530 \\
\textbf{NMPC Test @ 25$^{\circ}$C}         &\textbf{1.33e+12}  & \textbf{1.65e+12} &\textbf{0.340}  & \textbf{0.304} & \textbf{1800} \\
\textbf{DRL Test @ 25$^{\circ}$C}         & \textbf{2.90e+12} & \textbf{4.86e+12} & \textbf{0.244} & \textbf{0.233} & \textbf{1800} \\
\hline
\end{tabular}
\end{table*}

\subsection{Parameter Estimation Analysis}
This section evaluates the robustness of the parameter identification process for the proposed DRL-based experimental design compared to NMPC and conventional test cycles. The comparison aims to highlight the information richness of the DRL-based experiments relative to the model-based controller and standard test procedures in parameter estimation.\par
The target parameters (\(k_p\) and \(k_n\)) are perturbed from their nominal values (\(k_p = 2.43 \times 10^{-9}\) and \(k_n = 1.85 \times 10^{-9}\)). A least-squares optimization problem is formulated to minimize the squared error between the E-ECM voltage output with nominal $V_k$ and perturbed parameter $\widetilde{V}_k$ values over the test period, which consists of \(N\) time steps. The optimization problem is expressed as follows:
\begin{equation}
\theta_i^\ast = \arg\min \sum_{k=0}^{N-1} \left( V_k \left( \theta_i, I_k \right) - \widetilde{V}_k \left( \widetilde{\theta}_i, I_k \right) \right)^2
\end{equation}
\par The goal of this procedure is to determine if the least squares optimization can accurately converge to the nominal parameter values. The process begins with perturbed initial values \(\widetilde{\theta}_i\), and the optimizer searches for parameter values that best match the synthetic voltage data, which are previously generated using the E-ECM model with the nominal values of the parameters. In this single parameter estimation process, ten initial perturbed values for \(k_p\) and \(k_n\) are considered, denoted as $\widetilde{\theta}_i$ $(i=k_p, k_n)$, and are evenly distributed within a specified range of these parameters ($k_p\in [3\times 10^{-12}, 2.4\times 10^{-9}]$, $k_n\in [2\times 10^{-11}, 1\times 10^{-8}]$). Each initial perturbed condition converges to an optimal converged parameter value, which is then compared to the nominal parameter value \(\theta_i\).\par
\begin{figure}[h!]
    \centering
    \includegraphics[angle=0, scale=0.55]{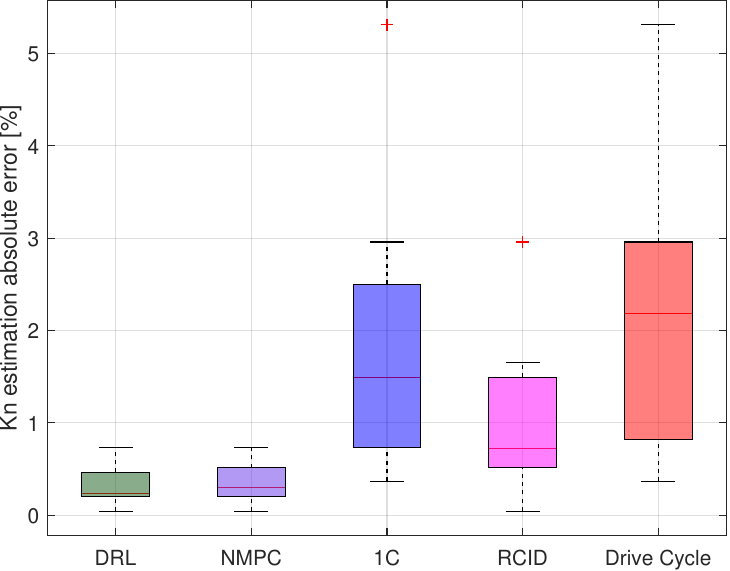}
    \caption{Parameter estimation absolute error distribution for anode rate constant $k_n$.}
    \label{fig:box_kn}
\end{figure}
\begin{figure}[h!]
    \centering
    \includegraphics[angle=0, scale=0.55]{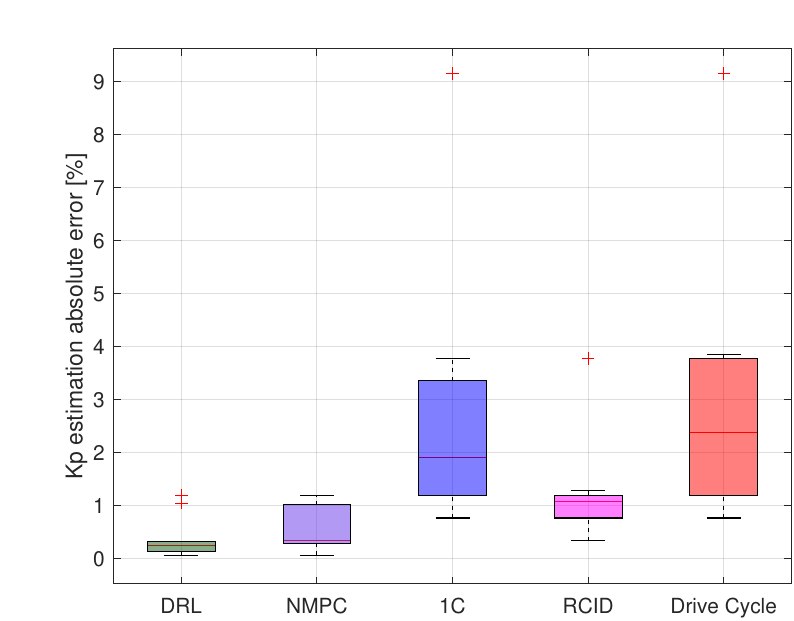}
    \caption{Parameter estimation absolute error distribution for cathode rate constant $k_p$.}
    \label{fig:box_kp}
\end{figure}

This analysis, as shown in Table \ref{table:1}, highlights significant performance differences among the tested profiles. The DRL-based test emerges as the most effective profile, demonstrating the highest average FI for both \( k_p \) and \( k_n \) and the lowest parameter estimation errors. Specifically, Fig. \ref{fig:box_kn} and Fig. \ref{fig:box_kp} illustrate that the DRL approach achieves the lowest median absolute estimation errors, 0.24\% for \( k_p \) and 0.23\% for \( k_n \), along with the least variability across test cases. This indicates not only superior accuracy but also the most consistent performance when compared to the NMPC approach and conventional tests. \par


It is worth noting that the NMPC approach performs more closely than the other conventional tests to the DRL approach in terms of parameter accuracy, particularly when assessing median values. However, the NMPC method incurs significantly higher computation times when controlling the input excitation (5 s vs. 5.2 ms on average for each time step) while acknowledging that the DRL approach requires a long training phase ($\sim$24 hrs) to learn the optimal policy. This increased computational burden of the NMPC design in online input excitation may restrict its practical applicability, despite achieving comparable parameter estimation accuracy. Furthermore, in the conventional tests, RCID test while effective for estimating \( k_n \) and \( k_p \), requires a much longer experimental duration. The drive cycle and 1C tests exhibit moderate performance, with higher estimation errors and less informative data, making it less effective overall. \par
The findings collectively demonstrate that the DRL-based input excitation is an efficient approach in optimal experimental design, combining high FI and low parameter estimation errors with a short experimental duration. 

\section{Conclusions and Future Work}
This paper demonstrates the potential of DRL in optimizing the experimental design for electrochemical parameter identification in LiBs. Comparative analysis with the NMPC method and conventional tests reveals that the DRL-based approach achieves higher FI and lower parameter estimation errors for anode and cathode rate constant parameters. Furthermore, the DRL approach shows a computational advantage, requiring less experimental time while delivering accurate parameter identification. In contrast, while NMPC also delivers accurate results, it suffers from higher computational complexity due to continuous optimization requirements. Future work will explore the integration of the learned DRL-policy to characterize an actual li-ion cell via hardware-in-the-loop and conduct parameter estimation with voltage measurements from the actual battery.
\addtolength{\textheight}{-12cm}   




\section*{ACKNOWLEDGMENT}
The authors would like to express their gratitude to Dr. Phillip Aquino of the Honda Research Institute for the insightful discussions and valuable feedback that contributed to the development of this research.

\section*{REFERENCES}

\begingroup
\renewcommand{\section}[2]{} 
\bibliographystyle{ieeetr}
\bibliography{mybibfile}
\endgroup

\end{document}